\documentclass[usenatbib]{mn2e}

\usepackage[dvips]{graphicx}
\usepackage{amssymb}
\usepackage{txfonts}
\usepackage{colordvi}

\newcommand{\xte}{{\textit{RXTE}}}

\newcommand{\msun}{{\rm M}_{\sun}}

\newcommand{\g}{$\gamma$}

\newbox\grsign \setbox\grsign=\hbox{$>$} \newdimen\grdimen \grdimen=\ht\grsign
\newbox\simlessbox \newbox\simgreatbox \newbox\simpropbox
\setbox\simgreatbox=\hbox{\raise.5ex\hbox{$>$}\llap
     {\lower.5ex\hbox{$\sim$}}}\ht1=\grdimen\dp1=0pt
\setbox\simlessbox=\hbox{\raise.5ex\hbox{$<$}\llap
     {\lower.5ex\hbox{$\sim$}}}\ht2=\grdimen\dp2=0pt
\setbox\simpropbox=\hbox{\raise.5ex\hbox{$\propto$}\llap
     {\lower.5ex\hbox{$\sim$}}}\ht2=\grdimen\dp2=0pt

\def\la{\mathrel{\copy\simlessbox}}

\title[The jet power of GRS 1915+105]{The jet kinetic power, distance and inclination of GRS 1915+105}

\author[A. A. Zdziarski]
{Andrzej A. Zdziarski\\
Centrum Astronomiczne im.\ M. Kopernika, Bartycka 18, PL-00-716 Warszawa, Poland\\
}

\date{Accepted 2014 July 26. Received 2014 July 20; in original form 2014 June 07}

\pagerange{\pageref{firstpage}--\pageref{lastpage}}
\pubyear{2014}

\begin{document}

\maketitle

\label{firstpage}

\begin{abstract}
We apply a recently developed technique of calculating the minimum jet kinetic power to the major mass ejections of the black-hole binary GRS 1915+105 observed in radio wavelengths in 1994 and 1997. We derive for them the distance-dependent minimum power, and the corresponding mass flow rate and the total energy and mass content. We find that a fast increase of the jet power with the increasing distance combined with the jet power estimates based on the bolometric luminosity imply the source distance is $\la 10$ kpc. If the jet in GRS 1915 contains ions, their bulk motion dominates the jet power, which was either neglected or not properly taken into account earlier. We also reconsider the parameters of the binary, and derive the current best estimates of the distance-dependent black-hole mass and the inclination based on existing measurements combined with the kinematic constraints from the mass ejections. We also find the measurement of the donor radius of Steeghs et al.\ implies the distance to the system of $\la 10$ kpc, in agreement with the estimate from the jet power. 
\end{abstract}

\section{Introduction}
\label{intro}

GRS 1915+105 is a Galactic black-hole binary with a number of highly interesting and unique properties. It is a long-lived transient, which outburst began in 1992 \citep*{cbl92} and is still lasting. It was the first Galactic source discovered to show superluminal motion (\citealt{mr94}, hereafter MR94). It is unusually highly variable in X-rays (e.g., \citealt{belloni00}), and it is one of the intrinsically brightest Galactic X-ray binaries, occasionally radiating above the Eddington limit \citep*{dwg04}. The minimum jet power in the superluminal ejection in 1994 was claimed to be highly super-Eddington \citep*{gliozzi99}. 

Here, we reconsider the minimum kinetic jet power, the mass flow rate, and the total energy and mass of the major mass ejections of GRS 1915+105 observed in 1994 (MR94) and in 1997 (\citealt{fender99}, hereafter F99). We use the method developed in \citet{zdz14} (hereafter Z14). It minimizes the kinetic jet power associated with both the jet internal energy content and the ion rest mass, correcting some earlier estimates. We note, in particular, that finding the minimum comoving energy, adding the associated ions and transforming to the observer frame does not give the minimum of the power (Z14). The jet power becomes extremely high close to the maximum kinematically allowed distance for a given ejection, but it also decreases very fast with the decreasing distance. 

Recently, the black-hole mass and distance to GRS 1915+105 were measured by \citet{steeghs13} (hereafter S13). We use their results to derive a distance-dependent estimate of the black-hole mass. We also use their measurement of the donor radius to derive a constraint on the distance. 

\section{The parameters of GRS 1915+105}
\label{1915}

The jet inclination, $i$, and the velocity, $\beta_{\rm j}$, can be measured using the angular velocities of the approaching and receding condensations, $\mu_{\rm a}$ and $\mu_{\rm r}$, respectively, under the assumption that the ejections were symmetric (MR94). We have the relationships,
\begin{equation}
i=\arctan{2\mu_{\rm a}\mu_{\rm r} D\over (\mu_{\rm a}-\mu_{\rm r})c},\quad
\Gamma_{\rm j}={\mu_{\rm a}+\mu_{\rm r}\over 2\left[\mu_{\rm a}\mu_{\rm r}\left(1-{D^2\over c^2}\mu_{\rm a}\mu_{\rm r}\right)\right]^{1/2}},
\label{incl_gamma}
\end{equation}
where $D$ is the source distance and $\Gamma_{\rm j}$ is the jet Lorentz factor. [Note that $\sin(\arctan{x})=x/(1+x^2)^{1/2}$, which can be used in equations (\ref{q_r2}--\ref{mbh}) below for a given measurement of $\mu$.] From $\Gamma_{\rm j}(D)$, we see there is the maximum possible distance, $D_{\rm max}=c(\mu_{\rm a}\mu_{\rm r})^{-1/2}$, corresponding to $\Gamma_{\rm j}\rightarrow \infty$. The Doppler factors, $\delta_{\rm a,r}=\Gamma_{\rm j}^{-1}(1\mp \beta_{\rm j}\cos i)^{-1}$, of the approaching and receding ejecta are,
\begin{equation}
\delta_{\rm a}=\left[{\mu_{\rm a}\over \mu_{\rm r}}\left(1-{D^2\over c^2}\mu_{\rm a}\mu_{\rm r}\right)\right]^{1/2}\!,\quad \delta_{\rm r}=\delta_{\rm a}{\mu_{\rm r}\over \mu_{\rm a}}=
\left[{\mu_{\rm r}\over \mu_{\rm a}}\left(1-{D^2\over c^2}\mu_{\rm a}\mu_{\rm r}\right)\right]^{1/2}\!,
\label{delta}
\end{equation}
respectively. We then calculate the source-frame spectral luminosity from the approaching blob at $\nu/\delta_{\rm a}$ as 
\begin{equation}
L_{\rm j}(\nu/\delta_{\rm a})=4\upi D^2 F_{\rm a}(\nu)\delta_{\rm a}^{-3},
\label{lum}
\end{equation}
where $F_{\rm a}(\nu)$ is the measured flux of the approaching blob at $\nu$. We note that MR94 and F99 found that the observed ratios between the fluxes of the approaching and receding ejecta are somewhat lower than the ratio, $(\delta_{\rm a}/\delta_{\rm r})^{3+\alpha}$, where $\alpha$ is the energy spectral index ($F(\nu)\propto\nu^{-\alpha}$), implied by equation (\ref{lum}). Then, based on the results of MR94 only, \citet{aa99} proposed that the ejections were not intrinsically symmetric, which could explain the lower observed flux ratio. This seems not confirmed by the results of F99, who also found a lower flux ratio. The ratio could be lower if the emission were from continuous jet and counterjet \citep{sikora97}. However, the radio images, showing individual blobs, of MR94 and F99 appear not to be compatible with that. We will use equation (\ref{lum}), bearing in mind this issue as a caveat. Since the obtained values of $\delta_{\rm a}$ are close to unity for the values of $D$ we determine, a possible error due to this assumption is insignificant. Hereafter, we use the standard propagation of errors to calculate the uncertainties on the obtained quantities. 

\begin{figure}
\centerline{\includegraphics[width=6.2cm]{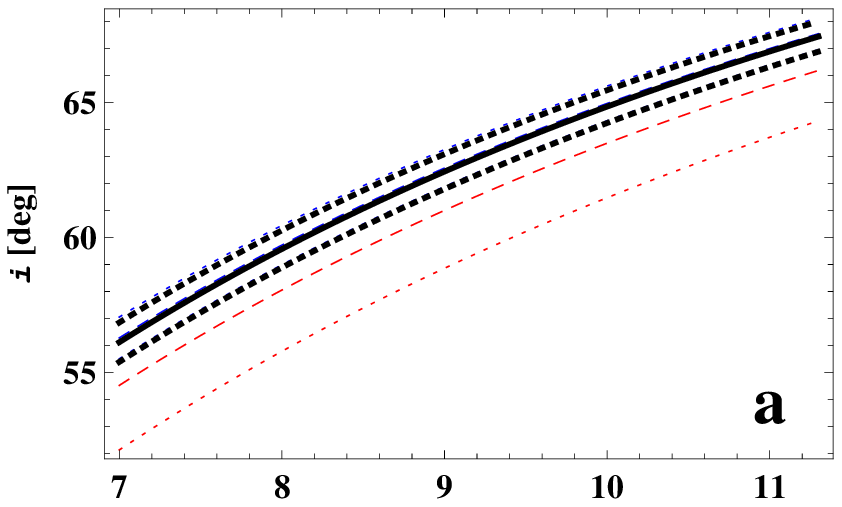}}
\centerline{\includegraphics[width=6.2cm]{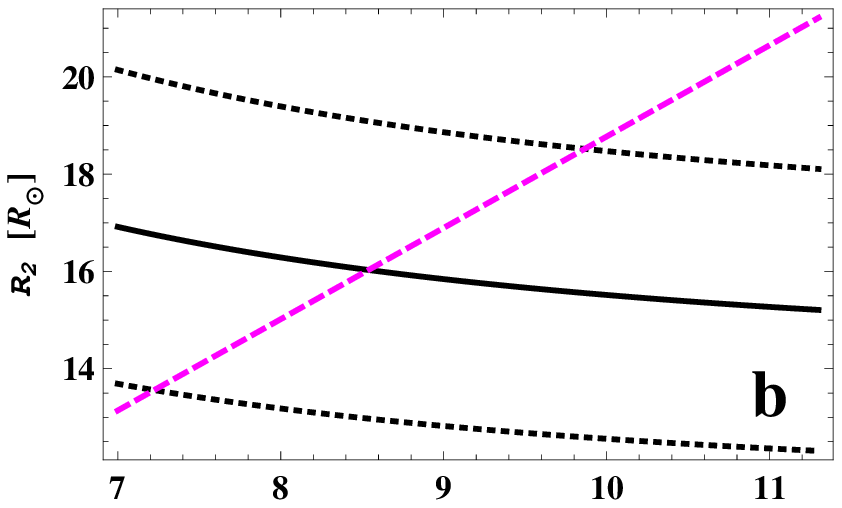}}
\centerline{\includegraphics[width=6.2cm]{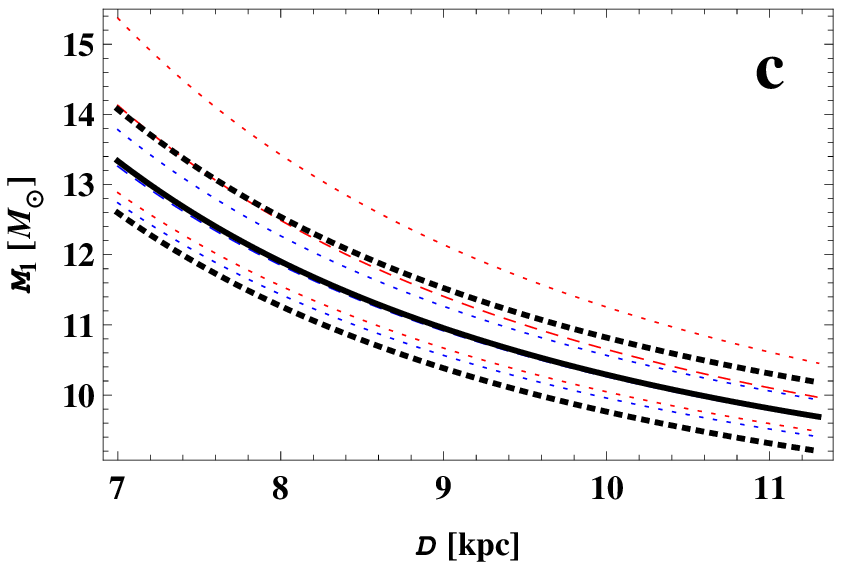}}
\caption{(a) The dependence of the inclination on the distance. The thin blue and red curves correspond to the 1994 March and 1997 October data, respectively, with the dashed and dotted curves corresponding to the values and the uncertainties, respectively. The heavy solid black curve gives the weighted average based on all the available data, and the dotted black curves give the uncertainties. (b) The radius of the secondary (solid black curve) with the uncertainties given by the dotted curves. The magenta dashed line gives the minimum radius allowed by the angular diameter of the K star implied by its observed magnitude \citep{z05}. We see that that the maximum $D$ allowed by this criterion is $\simeq 10$ kpc. (c) The black-hole mass obtained using the mass function of S13. The meaning of the curves is the same as in the panel (a). 
}
\label{binary}
\end{figure}

We consider two major ejection events from GRS 1915+105 which began on 1994 March 19 (MR94) and 1997 October 29 (F99). Their measured angular velocities were $\mu_{\rm a}=17.5\pm 0.3$ mas d$^{-1}$ (as updated in \citealt{rm99}), and $23.6\pm 0.5$ mas d$^{-1}$, and $\mu_{\rm r}=9.0\pm 0.1$ mas d$^{-1}$, and $10.0\pm 0.5$ mas d$^{-1}$, respectively. In addition, three more events with both $\mu_{\rm a}$ and $\mu_{\rm r}$ measured are listed by \citet{rm99} (see also \citealt{millerjones07}). The implied values of the inclination are consistent with being the same for all five events, though the additional three have much larger uncertainties. In particular, the values implied by the radial velocities for 1994 March and 1997 October are $(67.0\degr\pm 0.6\degr,\, 65.6\degr\pm 1.9\degr)$, $(65.0\degr\pm 0.6 \degr, 63.5\degr\pm 2.0\degr)$ and $(62.6\degr\pm 0.7\degr,\, 61.0\degr\pm 2.1\degr)$, at $D=11$, 10 and 9 kpc, respectively. Note that the stated uncertainties on $\mu$, and thus on $i(D)$, are significantly lower for the former event than for the latter. 

S13 argued, based on the results of \citet{sm12} (who corrected a numerical error by a factor of 50 in the alignment time scale given by \citealt{maccarone02}), that the axes of the jet and the binary plane in GRS 1915+105 are likely aligned, which assumption is also commonly adopted in other estimates of the binary parameters of GRS 1915+105. Still, the possibility of a misalignment should be kept in mind, given the lack of full understanding of the alignment process. Assuming alignment, the jet is not precessing, and its inclination angle remains the same for all the observations. Thus, we calculate the weighted average of the inclination and its standard error, though we note that it is strongly dominated by the two major events. Fig.\ \ref{binary}(a) shows the values of $i$ and its uncertainty for the two major events separately and for the average as a function of the distance. 

We then consider the implications for the binary parameters. We use the recent results of S13, who obtained the orbital period of $P_{\rm orb}=33.85\pm 0.16$ d, the velocity semi-amplitude of $K_2=126\pm 1$ km s$^{-1}$, and the rotational broadening of the donor of $v\sin i=21\pm 4$ km s$^{-1}$. Assuming corotation and the donor filling the Roche lobe, we have the standard formulae for the mass ratio, $q\equiv M_2/M_1$ (e.g., \citealt{wh88}), and the radius of the donor, $R_2$, 
\begin{equation}
q^{1/3}(1+q)^{2/3}={3^{4/3} (v\sin i)\over 2 K_2} ,\quad
R_2= {P_{\rm orb} (v\sin i) \over 2\upi \sin i},
\label{q_r2}
\end{equation}
respectively, where the Roche lobe radius ratio for $q\ll 1$ of \citet{paczynski71} has been used. The equation for $q$ yields a cubic-polynomial solution, which gives $q=0.043\pm 0.023$ (S13). On the other hand, $R_2$ is a function of $i$, which we show in Fig.\ \ref{binary}(b).

The black hole mass is
\begin{equation}
M_1={P_{\rm orb} K_2^3(1+q)^2\over 2\upi G\sin^3 i},
\label{mbh}
\end{equation}
and $M_2=q M_1$. S13 used the inclination range of F99. As we note above, it is significantly less accurate than that of MR94. Also, S13 gave their mass estimate including the entire distance-related inclination uncertainty of F99. Fig.\ \ref{binary}(c) shows instead the range of $M_1$ as a function of the distance and using the weighted average of the inclination from the available mass-ejection events. 

We then consider the distance. The maximum kinematic distances for the 1994 and 1997 events are $D_{\rm max}\simeq 13.6\pm 0.1$ kpc, $11.3\pm 0.3$ kpc, respectively, and thus the latter gives the current upper limit. The distance implied by the radial systemic velocity is $10.4\pm 1.3$ kpc (S13). \citet{z05} considered the distance implied by the theoretical K-star surface brightness compared to the observed magnitude, see their equation (1). This yields the angular diameter of the secondary as $s\simeq 0.0175$--0.0220 mas. The corresponding physical radius of the secondary is $R_2=D s/2$. The minimum $R_2$ as a function of $D$ allowed by the minimum $s$ above is plotted by the magenta dashed line in Fig.\ \ref{binary}(b). We see that the distance compatible with this constraint is $\la 10$ kpc. Finally, the new parallax distance from radio VLBA measurements is $8.6^{+2.0}_{-1.6}$ kpc \citep{reid14}. Combining the above determinations, we find the likely distance range as $D\simeq 9$--10 kpc. This is somewhat less than the distance of 11 kpc preferred by \citet{z05} because now both the current measurement of $v\sin i$ of S13 is lower than the previous one by \citet{hg04} and the determination of the systemic-velocity distance by S13 is lower than the previous one by \citet*{greiner01}. Our current range of $D$ is still compatible with other observational constraints listed in \citet{z05}. Also, a value of $D$ significantly less than 11 kpc is implied by considering the jet kinetic power, as we show below.

The values of $M_1$ for our preferred values of $D$ are then somewhat higher than $10.1\pm 0.6\,\msun$ given as the best estimate by S13, see Fig.\ \ref{binary}(c). The reason is simply our values of $D$ are lower than 10--12 kpc used by S13 to obtain their range of $M_1$. 

\section{The jet power of GRS 1915+105}
\label{power}

\begin{figure}
\centerline{\includegraphics[width=6.3cm]{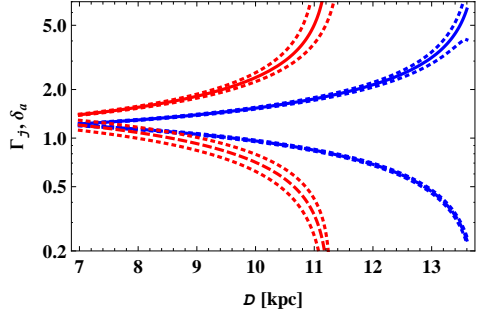}}
\caption{The dependencies of $\Gamma_{\rm j}$ (solid curves) and $\delta_{\rm a}$ (dashed curves) on $D$. The blue and red curves are for the 1994 (MR94; \citealt{rm99}) and 1997 (F99) flares. The dotted curves give the ranges of the uncertainties. 
}
\label{gamma_delta}
\end{figure}

\begin{figure}
\centerline{\includegraphics[width=4.8cm]{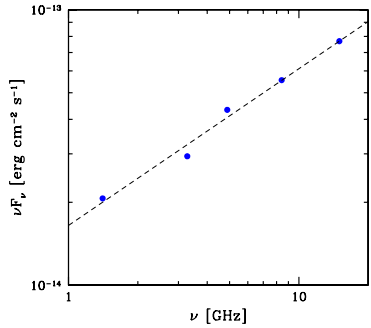}} 
\caption{The $\nu F_\nu$ radio spectrum of GRS 1915+105 observed on 1994 March 24 \citep{rodriguez95}. The spectrum is fitted by a power law ($F_\nu \propto \nu^{-\alpha}$) with $\alpha\simeq 0.43$, shown by the dashed line.
} \label{sp1915}
\end{figure}

MR94 gave rather accurate estimates of $\Gamma_{\rm j}$ and $i$ for their estimated $D=12.5$ kpc, but, as pointed out by \citet{fender03}, $\Gamma_{\rm j}$ strongly depends on $D$. This is illustrated in Fig.\ \ref{gamma_delta}, which shows the dependencies of $\Gamma_{\rm j}$ (increasing with $D$) for the best-fit values of the radial velocities of the two data sets. Fig.\ \ref{gamma_delta} also shows the dependencies of the Doppler factor $\delta_{\rm a}$, decreasing with $D$. 

\begin{figure}
\centerline{\includegraphics[height=5.5cm]{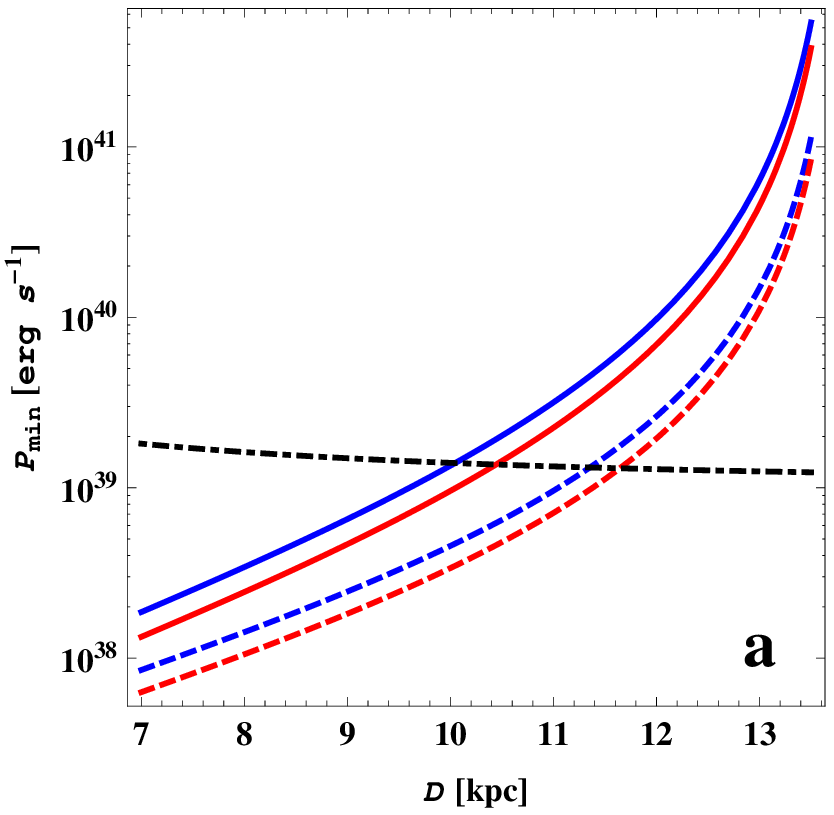}}
\centerline{\includegraphics[height=5.5cm]{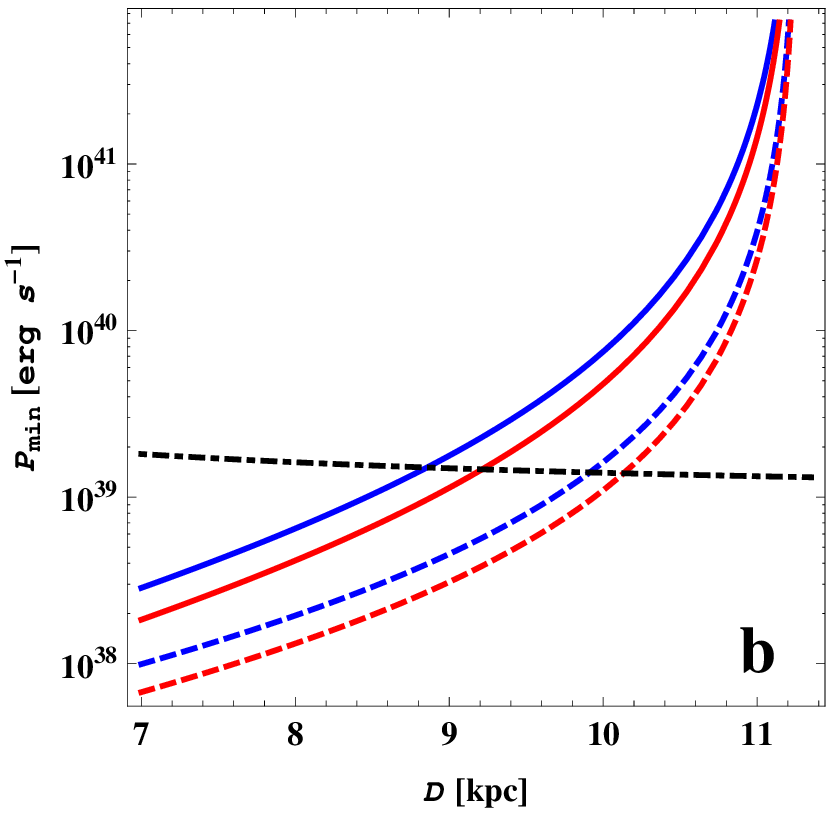}}
\caption{The minimum total jet+counterjet power as a function of the distance to GRS 1915+105 for (a) the 1994 and (b) 1997 ejections. The solid and dashed curves correspond to an electron-ion and an e$^\pm$ pair jet, respectively. The lower (red) and upper (blue) curves correspond to the cases of the emitting region in the jet being a uniform sphere or a spherical region with Gaussian distributions of the densities, respectively. The spread between those two models reflects systematic uncertainties of the modelling. The dot-dashed line give the Eddington luminosity, $L_{\rm E}$, at our distant-dependent estimate of the black hole mass, equation (\ref{mbh}).  
} \label{power1915}
\end{figure}

\begin{figure}
\centerline{\includegraphics[width=5.3cm]{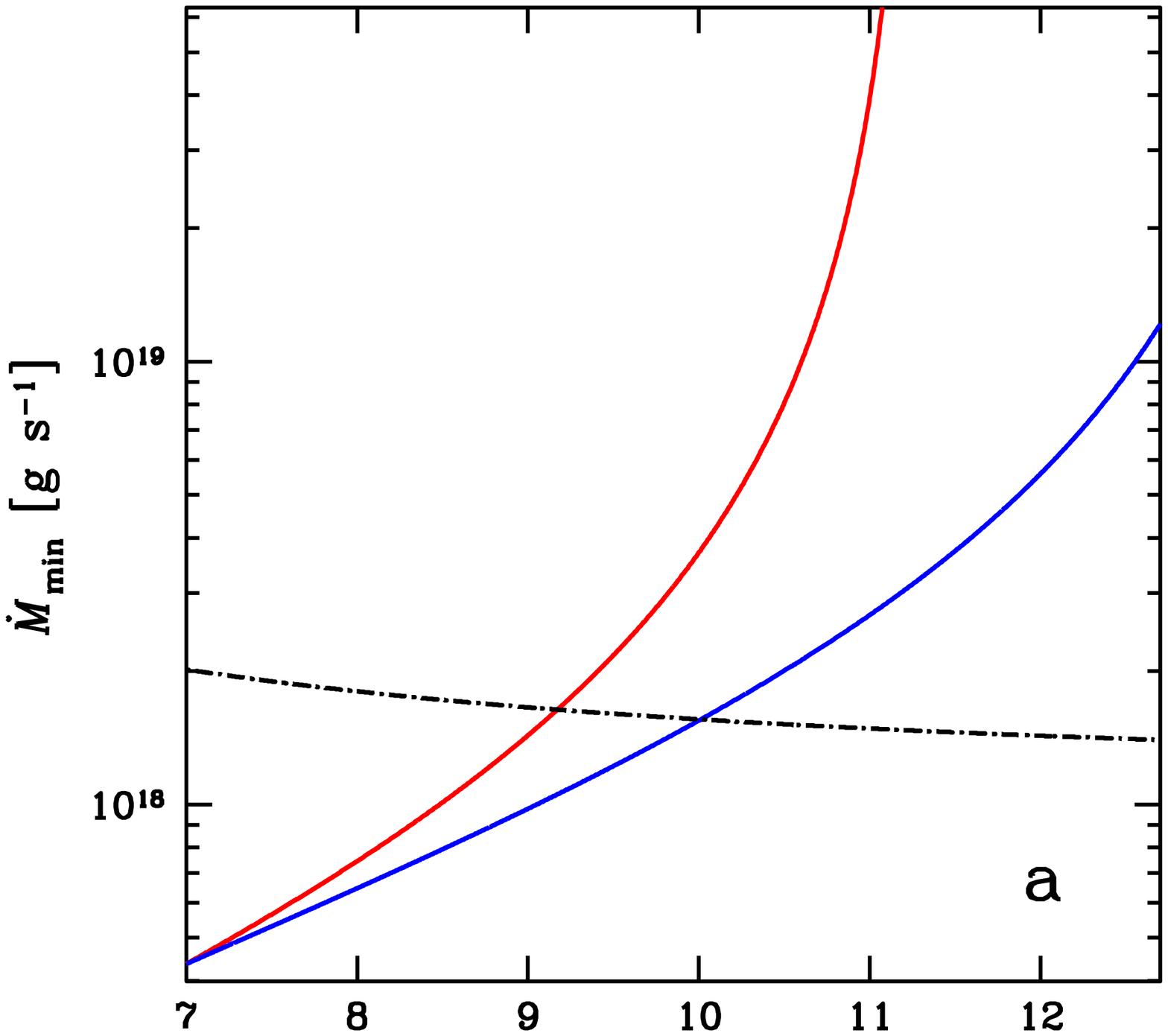}}
\centerline{\includegraphics[width=5.3cm]{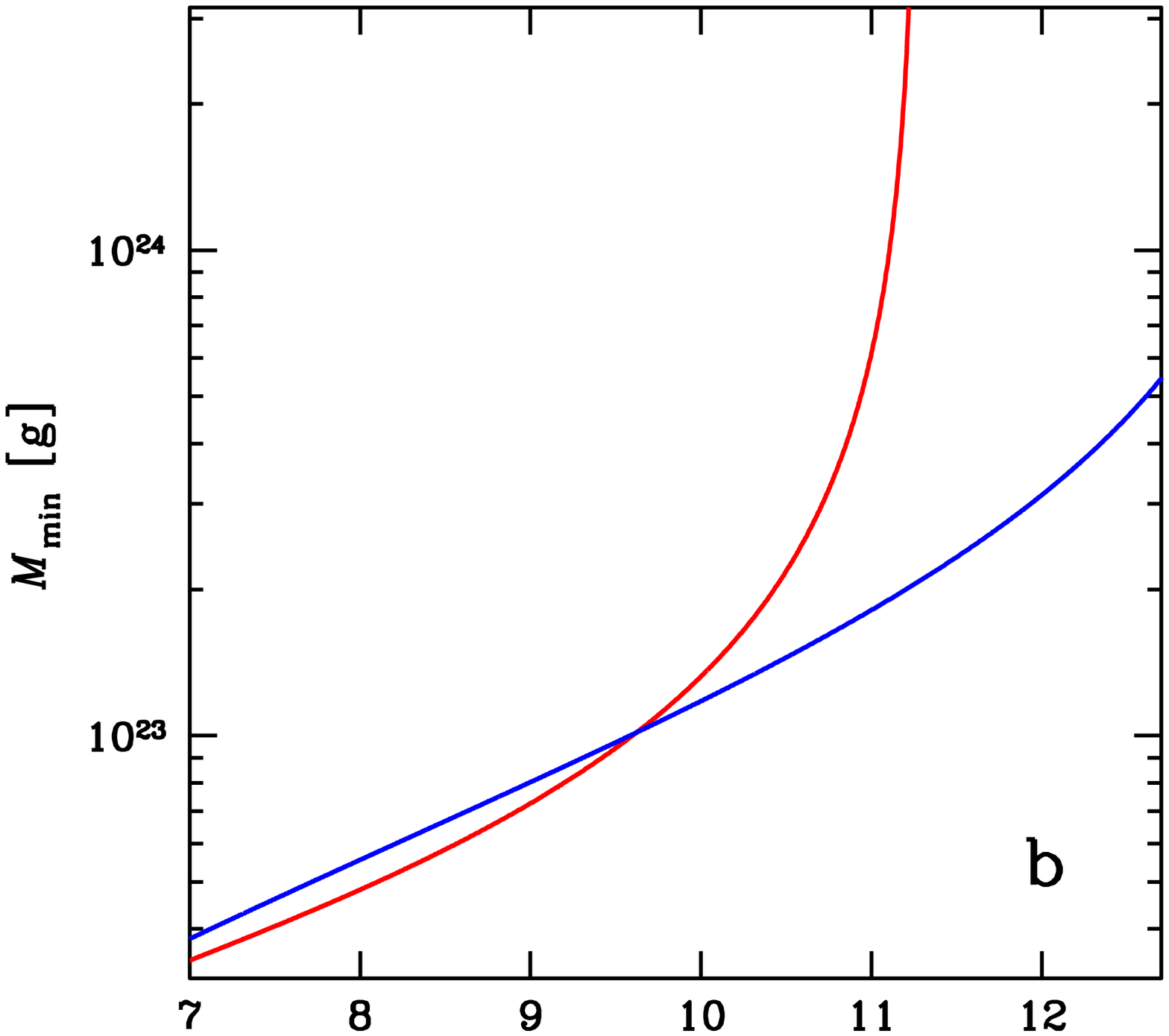}}
\centerline{\includegraphics[width=5.3cm]{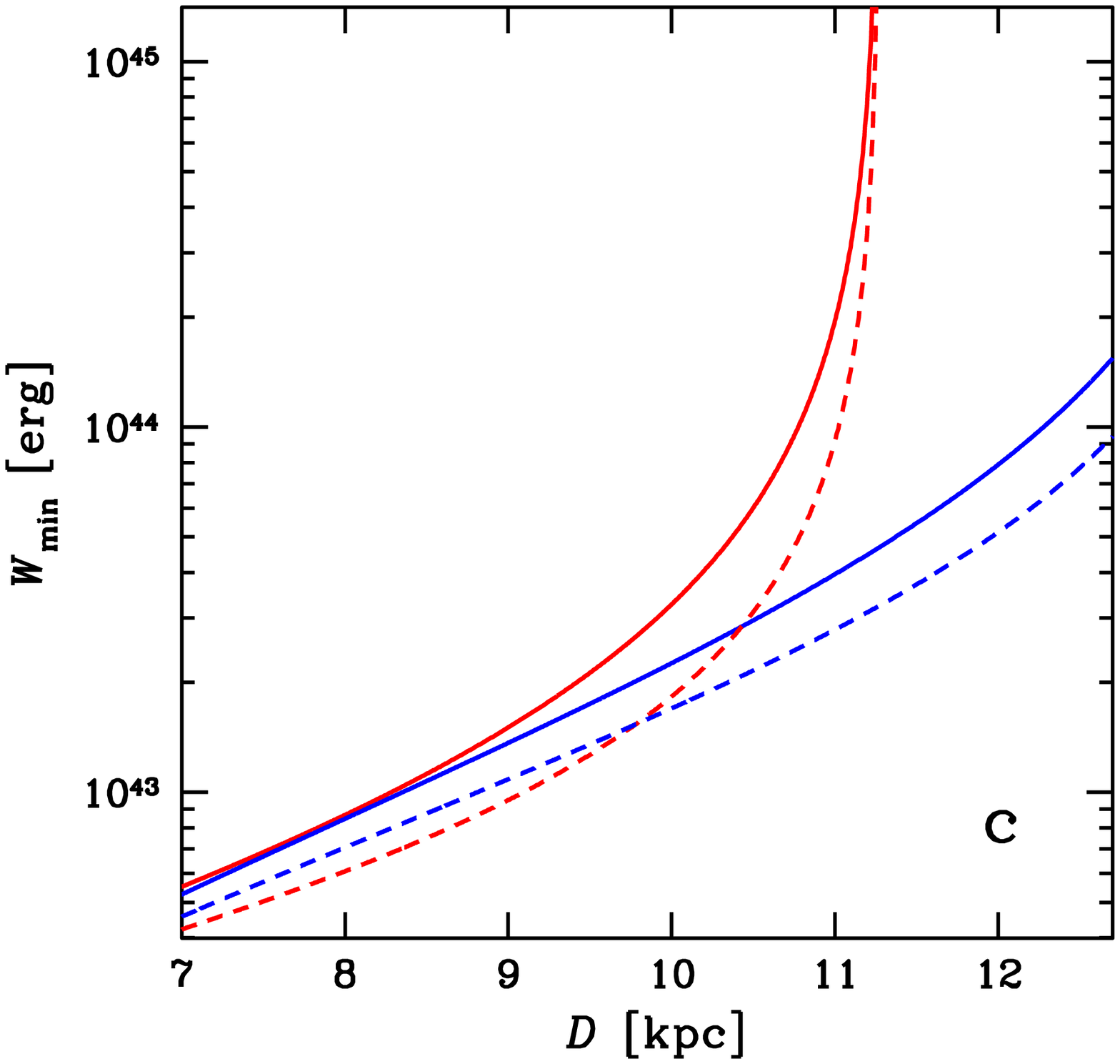}}
\caption{The dependencies of the minimum $\dot M$, $M$ and $W$ on $D$. The Gaussian model is shown for the 1994 (blue curves) and 1997 flare (red curves). (a) The minimum $\dot M$ for both jet and the counterjet. The dot-dashed curve corresponds to $\dot M_{\rm E}\equiv L_{\rm E}/c^2$ for the distance-dependent black-hole mass estimate; the accretion rate at the accretion luminosity $=L_{\rm E}$ is $\dot M_{\rm E}$ divided by the efficiency, $\sim 0.1$. (b) The total ion mass in both the approaching and receding ejection at the minimum jet power. The corresponding values for a pure e$^\pm$-pair jet are $\sim 2000$ times lower. (c) The total energy content in both particles and field corresponding to the minimum jet power. The solid and dashed curves correspond to electron-ion and pair plasmas, respectively.}
\label{mass}
\end{figure}

The measured peak flux of the 1994 flare was 660 mJy at $\nu= 8.4$ GHz on 1994 March 24 \citep{rodriguez95}. The observed 1.4--15 GHz spectrum is well fitted by a power law with the spectral index of $\alpha\simeq 0.43$ [$F(\nu)\propto \nu^{-\alpha}$], see Fig.\ \ref{sp1915}, corresponding to the electron power-law index of $p\simeq 1.86$. \citet{rodriguez95} give the total fluxes, but the flux of the approaching component dominates (MR94), and thus we use the total values. The measured peak flux of the 1997 flare was 550 mJy at $\nu= 2.3$ GHz on 1997 October 29 (F99). The data for that day are at 2.3 GHz, 8.3 GHz and 15 GHz. The 2.3--8.3 GHz spectrum has $\alpha\simeq 0.8$ (F99), though this slope under-predicts (at $\simeq$120 mJy) the average 15 GHz peak flux on that day, see fig.\ 4 of F99, which would imply a lower average $\alpha$. That flux also shows strong QPOs, implying a sequence of new ejections. Still, we adopt $\alpha=0.8$ for consistency with F99, corresponding to $p=2.6$.

We assume the spectrum is emitted from $\nu_{\rm min}=0.5$ GHz to $\nu_{\rm max}=50$ GHz (observer frame), consistent with the data. This photon energy range spans two decades, corresponding to one decade of the electron Lorentz factor. This is a conservative assumption, which yields good estimates of the {\it minimum\/} jet power. We assume that the only electrons, or e$^\pm$, are those in in a relativistic power-law distribution yielding the synchrotron power law from $\nu_{\rm min}$ to $\nu_{\rm max}$. We consider two cases of the jet composition, one with cold ions and the H abundance of $X=0.7$, and one with pure pair plasma. 

We adopt the angular source size (perpendicular) at the measured peak of the 1994 flare of 20 mas (MR94), which corresponds to the physical radius of $r_{\rm j}=1.5\times 10^{15}(D/10\,{\rm kpc})$ cm. For the 1997 flare, we adopt the size estimate of F99 of $r_{\rm j}=1.3\times 10^{15}$ cm independent of $D$ based on the jet rise time. In the calculations, we consider either a uniform spherical source, or one with smooth Gaussian density distributions, see Z14. We assume the blob has a spherical symmetry in its frame and thus we do not apply the relativistic transformation of the length along the jet. 

We use then the method minimizing the jet power for a given optically-thin synchrotron frequency range, see Z14. We use equations (\ref{incl_gamma}--\ref{lum}) above to determine the jet-frame $\Gamma_{\rm j}$ and $L_{\rm j}(\nu)$ as functions of the distance. The behaviour of $\delta_{\rm a}$ and $\Gamma_{\rm j}$ shown in Fig.\ \ref{gamma_delta} leads to a fast decline of $P_{\rm min}$ with decreasing $D$. 

At $D=12.5$ kpc, we obtain the the minimum jet power of $P_{\rm min}\simeq 1.5\times 10^{40}$ erg s$^{-1}$ of the 1994 flare for the uniform model, which is a factor of $\simeq 4$ lower than the corresponding estimate by \citet{gliozzi99} (multiplying their value by two to include the counterjet). The main cause of the difference is our more limited range of the electron Lorentz factors, which is $\gamma\in [44, 440]$, compared to $[1, 10^3$ of them. Also, our definition of the jet power is different, see Z14, and the method different, as they assumed a fixed frequency-integrated synchrotron luminosity. Taking into account those differences, our results are fully consistent with those of \citet{gliozzi99}. They claimed the 1994 flare had to be strongly super-Eddington. However, as we discuss above, the current limit on $D$ is $11.3\pm 0.3$ kpc (F99), and $P_{\rm min}$ decreases rather fast with decreasing $D$, which we show below. Thus, their results are no longer applicable to GRS 1915+105.

We then compare our results for the 1997 flare with those of F99. At $D=11$ kpc, we obtain $P_{\rm min}\simeq 1.4\times 10^{41}$ erg s$^{-1}$ with the uniform model, which is a factor of $>30$ higher than the corresponding estimate of F99 (multiplying the power given in their table 3 by 2, see their section 4.1.3). One reason for the discrepancy is their method being incorrect, as discussed in Z14. Furthermore, F99 divided the energy content in electrons by the flare rise time, which gives the power averaged over that time rather than the instantaneous power at the moment of the measurement of the peak synchrotron flux.

Figs.\ \ref{power1915}(a) and (b) show $P_{\rm min}(D)$ for the 1994 and 1997 flares, respectively, for the uniform and Gaussian models with the ionic and e$^\pm$ compositions (Z14), comparing them with the Eddington luminosity at our estimated $D$-dependent black-hole mass estimate. We see that $P_{\rm min}$ for the pair model is a factor of a few less than for the ionic one. However, as argued, e.g., by \citet{heinz08} and \citet{diaztrigo13} (though, as a caveat, see \citealt{neilsen14}), jets in X-ray binaries likely contain hadrons rather than pure pairs. The minimum jet+counterjet power and the corresponding magnetic field strength using the Gaussian model (used hereafter) for the 1994 flare at 10 kpc are $P_{\rm min}\simeq 1.3\times 10^{39}$ erg s$^{-1}$, $B_{\rm min}\simeq 0.11$ G for the cosmic abundances and $P_{\rm min}\simeq 4.6\times 10^{38}$ erg s$^{-1}$, $B_{\rm min}\simeq 0.07$ G for e$^\pm$ pairs. For the 1997 flare at 10 kpc, we obtain $P_{\rm min}\simeq 7.5\times 10^{39}$ erg s$^{-1}$, $B_{\rm min}\simeq 0.17$ G for the cosmic abundances and $P_{\rm min}\simeq 1.6\times 10^{39}$ erg s$^{-1}$, $B_{\rm min}\simeq 0.09$ G for e$^\pm$ pairs. At the cosmic abundances, we find the ratio of the jet power in electrons to that in ions of $P_{\rm e} /P_{\rm i}\simeq 0.18$, 0.08 for the 1994 and 1997 flare, respectively. These estimates are thus insensitive to the assumed source radius, which is the case for the ionic power dominating (Z14). Indeed, the values of the critical radius below which the solution becomes independent of $r_{\rm j}$ (Z14) are $r_{\rm cr}\simeq 4.4\times 10^{16}$ cm, $\simeq 2.2\times 10^{17}$ cm for the 1994 and 1997 flares, respectively, which are $\gg$ our values of $r_{\rm j}$. We note that a possible clumpiness of the jet material, as quantified by the clump volume filling factor, $f$, can reduce the minimum jet power by a factor of $f^{1/3}$ in the case of the dominant ionic power, and $f^{3/7}$ for the case of pair plasma (Z14). If, e.g., $f\sim 0.01$, the minimum power can be reduced by a factor of several.

Fig.\ \ref{power1915} shows that the kinetic jet power critically depends on the actual distance to GRS 1915+105. We can compare it to the bolometric luminosity of GRS 1915+105, dominated by accretion. The results of \citet{allen06} and \citet{mh07} show that the jet kinetic power in AGNs is strongly correlated with both the accretion rate and the bolometric luminosity, $L_{\rm bol}$. In particular, \citet{mh07} find the correlation of $\lg (P_{\rm j}/L_{\rm E}) = (0.49 \pm 0.07) \lg(L_{\rm bol}/L_{\rm E}) - 0.78 \pm 0.36$, where $L_{\rm E}$ is the Eddington luminosity. Thus, we expect $P_{\rm j}\la L_{\rm bol}$ at $L_{\rm bol}\sim L_{\rm E}$. Furthermore, results of \citet*{mhd03} and \citet{hs03} show that the same underlying relationship between jets and accretion flow is present in both AGNs and black-hole binaries. Indeed, the jet power of Cyg X-1 estimated by \citet{gallo05} satisfies the correlation of \citet{mh07}. Furthermore, the jet power of Cyg X-3 in its bright state has been estimated based on a model of its \g-ray emission and found $<L_{\rm bol}$, consistent with the above correlation \citep{z12}. The \xte/ASM monitoring shows the X-ray state during the 1997 event (F99) was among the brightest ones observed from this source, which implies that its maximum bolometric luminosity was then moderately super-Eddington, $\la 2 L_{\rm E}$ \citep{dwg04}. Thus, based on the above considerations, we may expect (but not prove) the jet kinetic power to be approximately sub-Eddington. Fig.\ \ref{power1915} shows that the 1994 and 1997 jets are sub-Eddington provided $D\la 10.1$ kpc and 8.8 kpc, respectively, using the ionic model with the Gaussian density distribution. In the case of pair plasma, the respective jets are sub-Eddington for $D\la 11.4$ kpc and 9.9 kpc.

The minimum ion mass flow rate, $\dot M_{\rm min}$, for the Gaussian model are shown in Fig.\ \ref{mass}(a). We see that if $D\la 10$ kpc, $\dot M_{\rm min}\la L_{\rm E}/c^2$. The corresponding masses for the ionic model (in the jet frame) are shown in Fig.\ \ref{mass}(b). In the case of the pair model, both the minimum $\dot M$ and $M$ are lower by $\sim \mu_{\rm i}m_{\rm p}/m_{\rm e}\sim 2000$. Fig.\ \ref{mass}(c) shows the total energy content corresponding to the minimum jet power for both ionic and pair cases. 

The ejecta at the time of the 1994 measurements were optically thin to synchrotron self-absorption; the radial optical depth, $\tau_{\rm S}$, at $\nu_{\rm min}=0.5$ GHz is $\simeq 0.43$, and $\simeq 10^{-4}$ at $\nu=8.6$ GHz. (Similar values are found for the 1997 flare.) The peak 8.6-GHz flux, corresponding to $\tau\sim 1$, was thus achieved earlier than on 1994 March 24, and the minimum power at that time was higher. On the other hand, the blob expanded after March 24, which caused the electrons to cool adiabatically, and also reduced the magnetic field strength. The synchrotron luminosity should then decrease, and the fluxes in table 1 of MR94 are consistent with this. When we apply the minimum jet power method to those later stages, we find the minimum total power decreasing. On the other hand, there was no evidence for deceleration in those data (MR94; see also \citealt{millerjones07}), which implies that the power in ions (dominating the power in particles in our case) kept constant. Thus, the further evolution of the source had to be characterized by a decrease of the jet equipartition parameter of $P_B/(P_{\rm e}+P_{\rm i})$ ($\simeq 0.5$ for the minimum power at the initial observation). This shows that the actual jet power is likely to be higher than the minimum one.

\section{Conclusions}
\label{conclusions}

We have calculated the minimum jet power for the major mass ejection events of GRS 1915+105 in 1994 and 1997. We have taken into account the power component associated with the ion bulk motion, which we have found to dominate, and which was neglected in an earlier estimate. We have found the jet power becomes very high close to the maximum allowed kinematically for a given ejection event. On the other hand, it becomes much lower and in agreement with the average relationship between the jet power and the bolometric luminosity for accreting black holes at a distance $\la 10$ kpc. We also have calculated the mass flow rate in the jets and their total mass and energy content.  

We have also updated the distance estimate based on a theoretical relationship of the donor radius and its magnitude for K stars \citep{z05} using the $v\sin i$ measurement of S13. We obtained $D\la 10$ kpc, in full agreement with the estimate based on the jet power. In addition, we have presented a relatively accurate estimate of the black-hole mass as a function of the distance using the weighted average of the inclination from the observed mass ejections together with the optical measurements of S13.

\section*{ACKNOWLEDGMENTS}

I am grateful to P. Pjanka for detailed checking of the results of this work, F. Mirabel, J. Miko{\l}ajewska and M. Sikora for valuable discussions, the referee for valuable comments and suggestions, and L. F. Rodr{\'{\i}}guez for a clarification regarding the size estimates in MR94. This research has been supported in part by the Polish NCN grants 2012/04/M/ST9/00780 and 2013/10/M/ST9/00729.

\label{lastpage}

\end{document}